\documentstyle[preprint,aps,eqsecnum]{revtex}   

\newcommand{\beq}{\begin{equation}}
\newcommand{\eeq}{\end{equation}}
\newcommand{\beqa}{\begin{eqnarray}}
\newcommand{\eeqa}{\end{eqnarray}}

\newcommand{\dm}{\mbox{$\Delta M$}}

\newcommand{\dphi}{\mbox{$\delta \phi$}}
\def\sm{Standard Model}

\def\ie{{\it i.e.}}
\def\eg{{\it e.g.}}
\newcommand{\lpp}{\mbox{$\lambda^{\prime \prime}$}} 
\newcommand{\qt}{\mbox{$\tilde q$}}
\def\npb#1{Nucl.\ Phys.\ {\bf B #1}}
\def\plb#1{Phys.\ Lett.\ {\bf B #1}}
\def\prd#1{Phys.\ Rev.\ {\bf D #1}}
\def\prl#1{Phys.\ Rev.\ Lett. {\bf #1}}

\def\ijmpa#1{Int.\ J.\ Mod.\ Phys.\ {\bf A #1}}
\def\progtp#1{Prog.\ Th.\ Phys.\ {\bf #1}}

\begin{document}

\draft

{\tighten
\preprint{
\vbox{
      \hbox{SLAC-PUB-7351}
      \hbox{hep-ph/9612269}
      \hbox{December 1996} }}
\bigskip
\bigskip

\renewcommand{\thefootnote}{\fnsymbol{footnote}}

\title{$CP$ Asymmetries in $B$ Decays \\
with New Physics in Decay Amplitudes
\footnotetext{Research supported
by the Department of Energy under contract DE-AC03-76SF00515}}
\author{Yuval Grossman and  Mihir P. Worah}
\address{ \vbox{\vskip 0.truecm}
Stanford Linear Accelerator Center \\
        Stanford University, Stanford, CA 94309}

\maketitle

\begin{abstract}%
We make a systematic analysis of the effects of new physics in
the $B$ decay amplitudes on the $CP$ asymmetries in 
neutral $B$ decays.
Although these are expected to be smaller than new physics effects 
on the mixing 
amplitude, they are easier to probe in some cases.
The effects of new contributions to the mixing amplitude are felt 
universally across all decay modes, whereas the effects of new decay 
amplitudes could vary from mode to mode.
In particular the prediction that 
the $CP$ asymmetries in the $B_d$ decay modes with
$b\to c\bar c s$, $b \to c\bar c d$, $ b\to c \bar u d$ 
and $ b\to s \bar s s$ 
should all measure the same quantity ($\sin2\beta$ in the \sm)
could be violated. 
Since the above \sm\ prediction is very precise, new decay amplitudes
which are a few percent of the \sm\ amplitudes can be probed.
Three examples of models where measurable effects are allowed are
given: 
effective supersymmetry, models with enhanced chromomagnetic dipole
operators, and supersymmetry without $R$ parity.

\end{abstract}

} 

\newpage

\section{Introduction}

$CP$ violation has so far only been observed in the decays of neutral
$K$ mesons. It is one of the goals of the proposed $B$ factories
to find and study $CP$ violation in the decays of $B$
mesons, and thus elucidate the mechanisms by which $CP$
violation manifests itself in the low energy world. 
There is a commonly accepted \sm\ of $CP$ violation,
namely that it is a result of the one physical phase in the $3 \times 3$
Cabbibo Kobayashi Maskawa (CKM) matrix \cite{ckm}. 
This scenario has specific predictions for the magnitude as well as
patterns of $CP$ violation that will be observed in the $B$ meson
decays \cite{Yossi}. 
However, since there currently exists only one experimental measurement of
$CP$ violation, it is possible that the \sm\ explanation for it is
incorrect, or more likely that in addition to the one CKM phase, there
are additional $CP$ violating phases introduced by whatever new
physics lies beyond the \sm.

In the limit of one dominant decay amplitude,
the $CP$ violating asymmetries measured in the time dependent
decays of neutral $B$
mesons to $CP$ eigenstates depend only on the sum of the phase of the 
$B^0 - \bar B^0$ mixing amplitude and the phase of the decay amplitude.
Although the CKM matrix could have up to five large 
phases (only one of which is independent), 
we know experimentally that only two of these are large.
This is manifest in the 
Wolfenstein \cite{wolf} parameterization where 
to leading order these phases are in the 
two CKM matrix elements $V_{ub}~(\gamma)$ and $V_{td}~(\beta)$.
In principle, one can determine $\beta$ and $\gamma$ from the 
available data on $K$ and 
$B$ decays. However, given the large theoretical uncertainties in the input
parameters (\eg\ $B_K$, $f_B$) the size of these 
phases remains uncertain \cite{gnalb,burasrev}.
Thus, the currently allowed range for the $CP$ asymmetries measurements in 
$B_d$ decays is very large.
Based on these facts the only precise predictions concerning the
$CP$ asymmetries made by the \sm\ are the following:
\begin{description}
\item[$(i)$]
{The $CP$ asymmetries in all $B_d$ 
decays that do not involve direct $b \to u$ (or $b \to d$)
transitions have to be the same.}
\end{description}
This prediction holds for the $B_s$ system in an even stronger form
\begin{description}
\item[$(ii)$]
{The $CP$ asymmetries in all $B_s$ 
decays that do not involve direct $b \to u$ (or $b \to d$) 
transition not only have to be the same, but also approximately vanish.}
\end{description}
Thus, the best place to look for evidence of new $CP$ violating
physics is obviously the $B_s$ system \cite{Nir-Sil,yuvalBs}. 
The $B$ factories, however,
will initially take data at the $\Upsilon(4s)$ where 
only the $B_d$ can be studied.

New physics could in principle contribute to both the mixing matrix and
to the decay amplitudes. It is plausible that 
the new contributions to the mixing could
be of the same size as the \sm\ contribution since it 
is already a one-loop effect.
This is why most of the existing studies on the
effects of new physics on $CP$ violating $B$ meson decays have
concentrated on effects in the mixing matrix, and assume the decay
amplitudes are those in the \sm\ \cite{Yossi,other,dutta} 
(in \cite{dutta} a more general analysis was done where they allow for 
new contributions to the penguin dominated \sm\ decay amplitudes).
The distinguishing feature of new
physics in mixing matrices is that its effect is universal, \ie\ 
although it changes the magnitude of the asymmetries it
does not change the patterns predicted by the \sm. 
Thus, the best way to search for these effects
would be to compare the observed $CP$ asymmetry in a particular
decay mode with the asymmetry predicted in the \sm. 
This is straightforward for the leading $B_s$
decay modes where the \sm\ predicts vanishing $CP$ asymmetries.
However, due to the large uncertainties in the \sm\ predictions
for the $B_d$ decays, these new effects would have to be
large in order for us to distinguish them from the \sm.
A slightly more sensitive analysis involves looking for
inconsistencies between the measured angles and sides of the unitarity
triangle \cite{worah,gronau-london96}. In any case, 
the \sm\ prediction $(i)$ concerning $B_d$ decays
still holds.

In contrast, the effects of new physics in decay amplitudes are
manifestly non-universal, \ie\ they depend on the specific process and
decay channel under consideration. Experiments 
on different decay modes that would measure
the same $CP$ violating quantity in the absence of new contributions
to decay amplitudes, now actually measure different $CP$ violating
quantities. Thus, the \sm\ prediction $(i)$,
concerning $B_d$ decays, can be violated.
Even though the possibility of new physics in decay
amplitudes is more constrained than that in mixing amplitudes,
one could detect these smaller
effects by exploiting the fact that now one does not care about the
predicted value for some quantity, only that two experiments that
should measure the same quantity, in fact, do not.
It is this possibility that we wish to study in this paper.

The outline of the paper is as follows: in Sec. II 
we first discuss the
general effects that new physics in decay amplitudes can have.
We then undertake a detailed discussion of each possible decay
channel, and the uncertainties in the universality predictions
introduced within the \sm\ itself by sub leading effects. 
Sec. III contains a brief
study of models of new physics that could contain new $CP$ violating
decay amplitudes, and their expected size. 
We present our conclusions in Sec. IV.

\section{The Effects of New Decay Amplitudes}

\subsection{General Effects}

In this sub-section we display the well known formulae for the decays
of neutral $B$ mesons into $CP$ eigenstates \cite{Yossi}, 
and highlight the 
relevant features that are important when more than one decay
amplitude contribute to a particular process.

The time dependent $CP$ asymmetry for the decays of states that
were tagged as pure $B^0$ or $\bar B^0$ at production into $CP$
eigenstates is defined as
\beq
a_{f_{CP}}(t) \equiv \frac{\Gamma[B^0(t) \rightarrow f_{CP}] -
                           \Gamma[\bar B^0(t) \rightarrow f_{CP}]}
                           {\Gamma[B^0(t) \rightarrow f_{CP}] +
                           \Gamma[\bar B^0(t) \rightarrow f_{CP}]},
\eeq
and given by
\beq
a_{f_{CP}}(t) = \frac{(1-|\lambda|^2)\cos(\Delta Mt) - 
                      2 Im \lambda \sin(\Delta Mt)}
                     {1 + |\lambda|^2},
\label{acp}
\eeq
where $\Delta M$ is the mass difference between the two physical
states, and 
\beq
\lambda = \left(\sqrt{\frac{M_{12}^{*}-\frac{i}{2}\Gamma_{12}^{*}}
               {M_{12}-\frac{i}{2}\Gamma_{12}}}\right)
          \frac{\langle f_{CP} |{\cal H} | \bar B^0 \rangle}
              {\langle f_{CP} |{\cal H} | B^0 \rangle}
        = e^{-2i\phi_M}\frac{\bar A}{A},  
\label{lambda}
\eeq
where we have used the fact that $M_{12} \gg \Gamma_{12}$, to replace
the first fraction in Eq.~(\ref{lambda}) by $e^{-2i\phi_M}$, the phase
of $B-\bar B$ mixing.

We now consider the case where the decay
amplitude $A$ contains contributions from two terms with 
magnitudes $A_i$,
$CP$ violating phases $\phi_i$ and $CP$ conserving phases $\delta_i$
(in what follows it will be convenient to think of $A_1$ giving the
dominant \sm\ contribution, and $A_2$ giving the sub leading \sm\
contribution or the new physics contribution).
\beq
A = A_1e^{i\phi_1}e^{i\delta_1} + A_2e^{i\phi_2}e^{i\delta_2}, \qquad
\bar A = A_1e^{-i\phi_1}e^{i\delta_1} + A_2e^{-i\phi_2}e^{i\delta_2}.
\eeq
To first order in $r \equiv A_2/A_1$ 
Eq.~(\ref{acp}) reduces to \cite{Gronau}
\beqa
a_{f_{CP}}(t)&=&\displaystyle{
-[2r\sin(\phi_{12})\sin(\delta_{12})]
\cos(\Delta Mt)} \nonumber \\
& & \displaystyle{
-[\sin 2(\phi_M + \phi_1) + 2r\cos 2(\phi_M + \phi_1)
\sin(\phi_{12})\cos(\delta_{12})]\sin(\Delta Mt)},
\label{a2}
\eeqa
and we have defined 
$\phi_{12}=\phi_1-\phi_2$ and $\delta_{12}=\delta_1-\delta_2$.

In the case $r=0$ or $\phi_{12}=0$
one recovers the frequently studied case where
$a_{f_{CP}}$ cleanly measures the $CP$ violating quantity $\sin
2(\phi_M + \phi_1)$. In addition, if there is no new physics
contribution to the mixing matrix (or if it is in phase with the \sm\
contribution), $a_{f_{CP}}$ cleanly measures $CP$ violating phases in
the CKM matrix. 

If $r \ne 0$ and $\phi_{12}\ne 0$ 
we can consider 3 distinct scenarios:

$(a)$ Direct $CP$ violation. This occurs when
$\delta_{12} \ne 0$ and can be measured by a careful
study of the time dependence since it gives rise to a $\cos\Delta Mt$
term in addition to the $\sin \Delta Mt$ term. Such a scenario would
also give rise to $CP$ asymmetries in charged $B$ decays.

$(b)$ Different hadronic final states even with the same quark content
could get different relative corrections, \ie, two different processes
with the same $\phi_1$ and $\phi_2$, but different $r$.
For example the decays $B_d \rightarrow D^{+}D^{-}$ and 
$B_d \rightarrow \psi\rho$ both go through the same quark level process 
$b\rightarrow c\bar c d$ and at leading order the $CP$ asymmetries both 
measure the same angle $\beta$   
(we have assumed that a transversality analysis allows us to 
treat $\psi\rho$ as a $CP$ eigenstate \cite{Lipkin}).
However the relative correction due to the \sm\ penguins themselves is 
expected to be
different for the two cases since the matrix elements are different.
Effects of this kind are
hard to estimate, and we will not study them further.

$(c)$ Different quark level 
decay channels that measure the same phase when only one
amplitude contributes, can measure different phases if more than one
amplitude contributes, \ie\ two different processes with the same
$\phi_1$, but with different $r$ or $\phi_2$.

Case $(a)$ demands
a non-vanishing strong phase difference which is hard to estimate. 
In order to get a valuable information
from  Case $(b)$ we need better theoretical understanding of hadronic
matrix elements.
Thus, we feel that case $(c)$
is the most promising way to
search for new physics effects in decay amplitudes, and we concentrate
on it for the rest of the paper. To this end we 
concentrate on the $\sin\dm t$ term in Eq.~(\ref{a2}) rewriting it as
\beq
a_{CP}(t)=-\sin 2(\phi_0 + \dphi)\sin\dm t,
\label{aa}
\eeq
where $\phi_0$ is the phase predicted at leading order in the \sm, and
$\dphi$ is the correction to it. For small $r$, $\dphi \le r$. However
for $r > 1$, $\dphi$ can take any value. Thus, when we have $\dphi \ge
1$ it should be understood that its value is arbitrary.

\subsection{The Different Decay Channels}

There are 12 different 
hadronic decay channels for the $b$ quark: 8 of them are
charged current mediated
\beqa
(c1)~ b \to c \bar c s\,, \qquad (c2)~ b \to c \bar c d\,, \qquad &&
(c3)~ b \to c \bar u d\,, \qquad (c4)~ b \to c \bar u s\,, \nonumber \\
(c5)~ b \to u \bar c d\,, \qquad  (c6)~ b \to u \bar c s\,, \qquad &&
(c7)~ b \to u \bar u d\,, \qquad  (c8)~ b \to u \bar u s\,, 
\eeqa
and 4 are neutral current
\beq
(n1)~b \to s \bar s s\,, \qquad  (n2)~ b \to s \bar s d\,, \qquad 
(n3)~b \to s \bar d d\,, \qquad  (n4)~ b \to d \bar d d\,.
\eeq
If only one \sm\ decay amplitude dominates all
of these decay channels, \ie\ $r = 0$ in Eq.~(\ref{a2}),
then up to 
${\cal O}(\lambda^2)$ (where $\lambda \approx 0.22$ is the
expansion parameter in the Wolfenstein approximation), 
the $CP$ asymmetries in $B$ meson decays all measure one of the 4
phases, 
\beqa  \label{SMfour} &&
\alpha \equiv \arg\left(-{V_{td}V_{tb}^*\over
V_{ud}V_{ub}^*}\right), \qquad
\beta \equiv \arg\left(-{V_{cd}V_{cb}^*\over
V_{td}V_{tb}^*}\right),\nonumber \\ &&
\gamma \equiv \arg\left(-{V_{ud}V_{ub}^*\over
V_{cd}V_{cb}^*}\right), \qquad
\beta' \equiv \arg\left(-{V_{cs}V_{cb}^*\over
V_{ts}V_{tb}^*}\right) \simeq 0.
\eeqa 
This situation is nicely summarized, along with relevant decay modes
in Table 1 of \cite{helen-pdg}.
Note that $\beta'< 2.5 \times 10^{-2}$ is very small in the 
SM \cite{burasrev}, 
but in principle measurable. 
For our purpose, however, this small value is a sub-leading correction to
the clean SM prediction $(ii)$. 
We will study corrections to this idealized limit, as well as to the
$r=0$ limit, in the next sub-section.
We now discuss the
effects that new physics in $b$ quark decay amplitudes could have on
the predictions of Eq.~(\ref{SMfour}).

In the \sm\ the $CP$ asymmetries in the decay 
modes $(c1)~b \to  c \bar c s$ (\eg\ $B_d \to
\psi K_S$, $B_s \to D_s^+ D_s^-$), 
$(c2)~b \to  c \bar c d$ 
(\eg\ $B_d \to D^{+}D^{-}$, $B_s \to \psi K_S$), and 
$(c3)~ b \to c \bar u d$ 
(\eg\ $B_d \to D^0_{CP}\rho$, $B_s \to D^0_{CP} K_S$)
all measure the angle $\beta$ in $B_d$ decay
and $\beta'$ in $B_s$ decays.
[$(c5)~b \to u \bar c d$ acts as a correction to $(c3)$ and will be
addressed later].
In the presence of new contributions to the $B-\bar B$ mixing matrix,
the $CP$ asymmetries in these modes would no longer be measuring the
CKM angles $\beta$ and $\beta'$. However, they would all still measure
the same angles $(\beta + \delta_{m_d}, \beta' + \delta_{m_s})$, 
where $(\delta_{m_d},\delta_{m_s})$ are the new contributions to the 
$B_{(d,s)}-\bar B_{(d,s)}$ mixing phase.
In contrast, new contributions to the $b$ quark decay amplitudes 
could affect
each of these modes differently, and thus they would each
be measuring different $CP$ violating quantities. 

Several methods \cite{grlo} have been proposed
based on the fact that the two amplitudes 
$(c4)~b \to c \bar u s$ 
and $(c6)~b \to u \bar c s$ (\eg\ $
B_d \to D_{CP}K_S$,
$B_s \to D_{CP}\phi)$ are comparable in
size, and contribute dominantly to the $D^0$ or $\bar D^0$ parts of
$D_{CP}$ respectively to extract the quantity
\beq \label{gamone}
\arg(b \to c \bar u s) + \arg(c \to d \bar d u)
- \arg(b \to u \bar c s) - \arg(\bar c \to \bar d d \bar u)
\equiv \gamma
\eeq
This measurement of $\gamma$ is
manifestly independent of the $B-\bar B$ mixing phase%
\footnote{%
We emphasize that 
$CP$ asymmetries into final states that contain
$D_{CP}$ cannot be affected by possible new contributions to $D - \bar D$
mixing. One
identifies $D_{CP}$ by looking for $CP$ eigenstate decay products like
$K^+K^-$, $\pi\pi$ or $\pi K_S$. 
As $(\Delta \Gamma / \Gamma)_D$ is known to be tiny,
the mass eigenstates cannot be identified. 
The relevant quantity that enters in the
calculation of the $CP$ asymmetry is 
the $D$ meson decay
amplitude and not the $D-\bar D$ mixing amplitude. Thus, the only new
physics in the $D$ sector
that could affect the standard analysis are new contributions
to the $D$ decay amplitudes.}.

The mode $(c7)~ b \to u \bar u d$ (\eg\ $B_d \to \pi \pi$, $B_s \to \rho
K_s$) measures the angles $(\beta+\gamma, \beta'+
\gamma$) in the \sm. 
We can combine this measurement, with the phase 
$(\beta, \beta'$) measured in the $(c1)~b \to c \bar c s$ 
mode to get another determination of $\gamma$ that is
independent of the phase in the $B-\bar B$ mixing matrix 
\eg\ comparing
$a_{CP}(t)[B_d \rightarrow \psi K_S]$ to
$a_{CP}(t)[B_d \rightarrow \pi \pi]$ allows us to extract
\beq \label{gamtwo}
\arg(b \to c \bar c d) - \arg(b \to u \bar u d) \equiv \gamma .
\eeq
Since both of the above evaluations of $\gamma$,
Eqs. (\ref{gamone}) and (\ref{gamtwo}) are
manifestly independent of any phases in the neutral meson mixing
matrices, 
the only way they can differ is if there
are new contributions to the $B$ or $D$ meson decay amplitudes.

The remaining charged current decay mode $(c8)~b \to u \bar u s$ 
suffers from large theoretical uncertainty 
since the tree and penguin contributions are similar in magnitude
and we will not study it here.

For the neutral current modes 
we will first assume that the dominant \sm\ contribution is from a
penguin diagram with a top quark in the loop, and discuss corrections
to this later. Since these are loop
mediated processes even in the \sm, $CP$ asymmetries into final states
that can only be produced by flavor changing neutral current vertices
are likely to be fairly sensitive to the possibility of new physics in
the $B$ meson decay amplitudes. The modes $(n3)~b \to s
\bar d d$ and $(n4)~b \to d \bar d d$ however, result in $CP$
eigenstate final states that are the same as 
for the charged current modes
$(c8)~b \to u \bar u s$ and $(c7)~b \to u \bar u d$ respectively.
Hence they cannot be used to study $CP$ violation, but rather act 
as corrections to the charged current modes.

In the \sm\ the 
mode $(n1)~ b \to s \bar s s$, (\eg\
$B_d \to \phi K_S$, $B_s \to \phi \eta')$ measures
the angle $\beta$ or $0$ in $B_d$ and $B_s$ decays. We can once
again try and isolate new physics in the decay amplitudes by comparing
these measurements with the charged current measurements of $\beta$.
Finally, $(n2)~ b \to d \bar s s$, \eg\
($B_d \to K_SK_S$, $B_s \to \phi K_S)$ 
measures the angle $0$ and $\beta$ for \sm\ $B_d$ and $B_s$ decays.

\subsection{Standard Model Pollution}

In all of the preceding discussion, we have considered the idealized
case where only one \sm\ amplitude contributes to a particular decay
process and we worked to first order
in the Wolfenstein approximation.
We would now like to estimate the size of the sub-leading
\sm\ corrections to the above processes, which then allows us to
quantify how large the new physics effects have to be in order for
them to be probed, and what are
the most promising modes to study.

There is a \sm\ penguin contribution to $(c1)~b \to c \bar c s$.
However, as is well known, this contribution has the same phase as the
tree level contribution (up to corrections of order $\beta'$) and
hence $\dphi=0$ in Eq.~(\ref{aa}). Thus in the absence of new
contributions to decay amplitudes,
the decay $B_d \to \psi K_S$ 
cleanly measures the phase $\beta+\delta_{m_d}$ 
(where $\delta_{m_d}$
denotes any new contribution to the mixing phase). The mode $(c2)~b
\to c \bar c d$ also has a penguin correction in the \sm. However, 
in this case 
$\phi_{12}={\cal O}(1)$ and we estimate
the correction as \cite{gronau,gronau-london96}
\beq
\dphi_{SM}(b \to c\bar c d) \simeq
\frac{V_{tb}V_{td}^*}{V_{cb}V_{cd}^*}\frac{\alpha_s(m_b)}{12 \pi}
\log(m_b^2/m_t^2)\lesssim 0.1 ,
\eeq
where the upper bound is obtained for $|V_{td}| < 0.02$, $m_t=180$ GeV
and $\alpha_s(m_b)=0.2$. The mode $(c3)~ b \to c \bar u d$ does not get
penguin corrections, however there is a doubly Cabbibo suppressed tree
level correction coming from $(c5)~ b \to u\bar c d$. Thus $B_d \to
D_{CP} \rho$ gets a second contribution with different CKM elements. 
While in general $\dphi$ can be a function of hadronic matrix
elements, 
here we expect this dependence to be very weak \cite{koide}.
In the factorization approximation, the matrix elements  
of the leading
and sub-leading amplitude are identical, as are 
the final state rescattering effects.  
Moreover, both these cases get contributions from only
one electroweak diagram,
thus reducing
the possibility of complicated interference
patterns. We then estimate
\beq
\dphi_{SM}(b \to c\bar u d) =
\frac{V_{ub}V_{cd}^{*}}{V_{cb}V_{ud}^{*}} r_{FA}
\le 0.05 .
\eeq
where $r_{FA}$ is the ratio of matrix elements with $r_{FA}=1$ in the
factorization approximation.
We have used $|V_{ub}/V_{cb}| < 0.11$, and used what we believe is a 
reasonable limit for the matrix elements ratio,
$r_{FA}<2$, to obtain the upper bound. 

The technique proposed to extract $\gamma$
using the modes $(c4)~b \to c \bar u s$ and $(c6)~b \to u \bar c s$
is manifestly independent of any ``\sm\ pollution''.
Finally $(c7)~b \to u\bar u d$ suffers from
significant \sm\ penguin pollution, which we estimate as 
\cite{gronau,gronau-london96} 
\beq \label{pen}
\dphi_{SM}(b \to u \bar ud) \simeq
\frac{V_{tb}V_{td}^{*}}{V_{ub}V_{ud}^{*}}\frac{\alpha_s(m_b)}{12 \pi}
\log(m_b^2/m_t^2)\lesssim 0.4 ,
\eeq
where the upper bound is for $|V_{td}| < 0.02$, $|V_{ub}|>0.002$,
$m_t=180$ GeV and $\alpha_s(m_b)=0.2$. The effects of the \sm\ penguin can
be removed by an isospin analysis \cite{gron-lon2}. 
However, this technique would then also rotate away any 
new physics contributions to the glounic penguin operator.

For the neutral current modes $(n1)~b \to s \bar s s$ 
the sub-leading \sm\ contribution is in phase with the
dominant contribution.
However, in the absence of new decay amplitudes,
the $CP$ asymmetry in $B_d \to \phi K_S$ will measure the angle
$\beta-\beta'+\delta_{m_d}$ and, $\dphi_{SM} = \beta' \le 0.025$.
Another source of uncertainty comes from $SU(3)_{flavor}$ mixing.
The $\phi$ also contains a small part of $u \bar u$, and thus
$B_d \to \phi K_S$  can also be mediated via the tree level 
$b \to u \bar u s$
decay that has a different weak phase than the leading penguin diagram.
From the data \cite{PDG} we can conservatively estimate that 
this extra uncertainty is about $1\%$. 
Combining these two sources of uncertainty we conclude
\beq
\dphi_{SM}(b \to s \bar ss) \le 0.04.
\eeq
This uncertainty can be reduced once $\beta'$ is measured, using
\eg\ $B_s \to D_s^+ D_s^-$.

Finally, $(n2)~b \to d \bar s s$ suffers from an 
${\cal O}(30\%)$ correction due to \sm\ 
penguins with up and charm quarks \cite{fleischer}. 

In summary, the cleanest modes are $b \to c \bar cs$ 
and $b \to c \bar us$ since they are essentially free
of any sub-leading effects. The modes $b \to c \bar ud$ and 
$b \to s \bar ss$ suffer only small theoretical 
uncertainty, less than $0.05$.
For $b \to c \bar cd$ the uncertainty is larger, ${\cal O}(0.1)$,
and moreover cannot be estimated reliably since it depends on the
ratio of tree and penguin matrix elements. 
Finally, the $b \to u \bar ud$ and $b \to d \bar ss$ 
modes suffer from large uncertainties.

\section{Models}

In this section we discuss three models that could have
experimentally detectable effects on $B$ meson decay amplitudes, and
violate the \sm\ predictions $(i)$ and $(ii)$. 
We also discuss ways to
distinguish these models from each other.

{\bf $(a)$ Effective Supersymmetry:} 
This is a supersymmetric extension of the \sm\ that seeks to retain
the naturalness properties of supersymmetric theories, while avoiding
the use of family symmetries or ad-hoc supersymmetry breaking boundary
conditions that are required to solve the flavor problems generic 
to these models \cite{ckn,dine-kagan}. In this
model, the $\tilde t_L$, $\tilde b_L$, $\tilde t_R$ and the gauginos
are light (below 1 TeV), while the rest of the super-partners are
heavy ($\sim 20$ TeV). The bounds on the squark mixing angles in this
model can be found in \cite{ckn2}. Using the formulae in \cite{masiero}
we find that for $\tilde b_L$ and gluino masses in the $100
- 300$ GeV range, this model generates $b \to s q \bar q$ and $b \to d
q \bar q$ transition amplitudes 
via gluonic penguins that could be up to twice as large as the \sm\
gluonic penguins, and with an unknown phase. Thus this model could
result in significant deviations from the predicted patterns of $CP$
violation in the \sm. We estimate these corrections to be
\beqa
\dphi_A(b \to c \bar c s) \lesssim 0.1, \qquad &&
\dphi_A(b \to c \bar c d) \lesssim 0.2, \qquad 
\dphi_A(b \to u \bar u d) \lesssim 0.8, \qquad \nonumber \\
\dphi_A(b \to s \bar s s) \lesssim 1, \qquad &&
\dphi_A(b \to d \bar s s) \lesssim 1, \qquad 
\eeqa
%

{\bf $(b)$ Models with Enhanced Chromomagnetic Dipole Operators:}
These models have been proposed to explain the discrepancies between
the $B$ semi-leptonic branching ratio, the charm multiplicity in $B$
decays and the \sm\ prediction for these quantities.
These enhanced chromomagnetic dipole operators 
come from gluonic penguins that arise naturally in TeV
scale models of flavor physics \cite{kagan}. 
In order to explain the above discrepancies with the \sm, these models
have amplitudes for $b \to s g$ that are 
about 7 times larger than the
\sm\ amplitude. 
The $b \to s q\bar q$ transition in this model is dominated by the
dipole operator for $b \to sg$ through the chain $b \to s g^{*} \to
sq\bar q$.
This interferes with the \sm\ $b \to s q\bar q$ amplitude. 
For the $B \to X_s \phi$
the net result is 
that the new amplitudes can be up to 
a factor of two larger than the \sm\
penguins and with arbitrary phase \cite{kagan2}. It is thus
plausible that similar enhancements can be present in 
the exclusive
$b \to c \bar c s$ transitions as well.
In addition, $b \to d g$
can be as large as $b \to s g$. However in the \sm\ the $ b
\to d$ penguins are Cabbibo suppressed compared to the $b \to s$
penguins. Thus in this model the corrections to the $b \to d \bar q
q$ modes could be much larger than the corrections to the $b \to s
\bar q q$ modes.
In the explicit models that have been studied, the
relative corrections to the $b \to dg$ \sm\ amplitude are up to 3 times
larger than those to the \sm\ $b\to s g$ amplitude \cite{kagan2}.
We estimate the following corrections to the dominant \sm\ amplitudes 
\beqa
\dphi_B(b \to c \bar c s) \lesssim 0.1, \qquad && 
\dphi_B(b \to c \bar c d) \lesssim 0.6, \qquad
\dphi_B(b \to u \bar u d) \lesssim 1, \nonumber \\
\dphi_B(b \to s \bar s s) \lesssim 1, \qquad &&  
\dphi_B(b \to d \bar s s) \lesssim 1.
\eeqa
%

{\bf $(c)$ Supersymmetry without R-parity:}
Supersymmetric extensions of the \sm\ usually assume the existence of
a new symmetry called $R$-parity. However, phenomenologically viable
models have been constructed where $R$-parity is not conserved
\cite{rparity}. In the absence of $R$-parity, baryon and lepton number
violating terms are allowed in the superpotential. Here we assume that 
lepton number is conserved in order to avoid bounds from proton decay
and study the effects of possible baryon number violating terms. The
relevant terms in the superpotential are of the form
$\lpp_{ijk} \bar u_i \bar d_j \bar d_k$,
where antisymmetry under $SU(2)$ demands $j \ne k$. The tree-level
decay amplitudes induced by these couplings are
then given by
\beq
A(b \to u_i \bar u_j d_k) \approx {\lpp_{i3l} \lpp_{jkl} 
\over 2 m_{\qt}^2}, \qquad
A(b \to d_i \bar d_j d_k) \approx {\lpp_{l3j} \lpp_{lik} 
\over 2 m_{\qt}^2}.
\eeq
Note that due to the requirement $i \ne k$ in the neutral current
mode, the decay $ b\to s\bar s s$ will not be corrected.
If we use, $m_{\tilde q} \simeq M_W$ for the squark masses, and assume
that there are no significant cancellations between the (possibly several)
terms that contribute to a single decay, then the bounds for
the relevant coupling constants are 
\cite{CRS}
\beq \label{rareBf}
\lpp_{ibs} \lpp_{ids} \lesssim 5 \times 10^{-3} , \qquad
\lpp_{ibd} \lpp_{isd} \lesssim 4.1 \times 10^{-3}, \qquad
\lpp_{ubs} \lpp_{cds} \lesssim 2 \times 10^{-2}.
\eeq
(We have imposed the
last bound in Eq.~(\ref{rareBf}) by demanding that the new contribution
to the $B$ hadronic width be less than the contribution from the \sm\
$b \to c \bar u d$ decay mode).
These lead to the following corrections to the dominant \sm\
amplitudes 
\beqa
\dphi_C(b \to c \bar c s) \lesssim 0.1, \qquad &&
\dphi_C(b \to c \bar c d) \lesssim 0.6, \nonumber \\
\dphi_C(b \to c \bar u d) \lesssim 0.5, \qquad &&  
\dphi_C(b \to d \bar s s) \lesssim 1.
\eeqa
%

The observed pattern of $CP$ asymmetries can also
distinguish between different classes of new contributions to the $B$
decay amplitudes. Here we list a few examples:

\noindent $(1)$ In model $(a)$ the maximum allowable 
relative corrections to the $b\to s$
and the $b \to d$ \sm\ amplitudes are similar in size.
While in model $(b)$ the relative corrections to the
$b \to d$ amplitude can be much larger.

\noindent $(2)$ 
In both models $(a)$ and $(b)$, the neutral current decay $b \to s
\bar s s$ can get significant $[{\cal O}(1)]$ corrections. In model
$(c)$ however, this mode is essentially unmodified. 

\noindent $(3)$ 
The fact that the $b \to c \bar u d$ channel can be
significantly affected in model $(c)$
is in contrast with the other two models.
In those models the new decay amplitudes were penguin
induced, and required the up-type quarks in the final state to be a
flavor singlet ($c\bar c$ or $u\bar u$), thus giving no correction to 
the $b \to c \bar u d$ decay.

\section{discussion and Conclusions}

Table~1 summarizes the relevant decay modes with their \sm\ uncertainty,
and the expected deviation from the \sm\ prediction in 
the three models we gave as examples.
New physics can be probed 
by comparing two experiments that measure the same phase $\phi_0$ in
the \sm\ [see Eq. (\ref{aa})].
A signal of new physics will be if these two measurements differ by an
amount greater than
the \sm\ uncertainty (and the experimental sensitivity) \ie 
\beq
|\phi(B \to f_1) - \phi(B \to f_2)| >
\dphi_{SM}(B \to f_1) + \dphi_{SM}(B \to f_2).
\eeq
Where $\phi(B \to f)$ is the angle obtained
from the asymmetry measurement in the $B \to f$ decay.

The most promising way to look for new physics effects in decay
amplitudes is to compare all
the $B_d$ decay modes that measure $\beta$ in the \sm\ 
(and the $B_s$ decay modes that measure $\beta'$ in the \sm).
The theoretical uncertainties among all the decays considered
are at most ${\cal O}(10\%)$,
and they have relatively large rates.
The best mode is $B_d \to \Psi K_S$ which has a sizeable rate and negligible 
theoretical uncertainty. This mode should be the reference mode to which all
other measurements are compared.
The $b \to c \bar u d$ and $b \to s \bar s s$ modes are also 
theoretically very clean.
In both cases the conservative
upper bound on the theoretical uncertainty is less than $0.05$,
and can be reduced with more experimental data.
Moreover, the 
rates for the relevant hadronic states are ${\cal O}(10^{-5})$ 
which is not extremely small.
Thus, the two ``gold plated'' relations are 
\beq
|\phi(B_d \rightarrow \psi K_S) - \phi(B_d \rightarrow D_{CP} \rho)| < 0.05, 
\eeq
and
\beq
|\phi(B_d \rightarrow \psi K_S) - \phi(B_d \rightarrow \phi K_S)| < 0.04.
\eeq
Any deviation from these two relations will be a clear indication for new 
physics in decay amplitudes.

Although not as precise as the
previous predictions, looking for violations of the relation
\beq
|\phi(B_d \rightarrow \psi K_S) -
\phi(B_d \rightarrow D^+D^-)| < 0.1,
\eeq
is another important way to search for new physics in the $B$ decay
amplitudes. The advantage is that the relevant rates are rather large,
$BR(B_d \rightarrow D^+D^-) \approx 4 \times 10^{-4}$. However, the theoretical
uncertainty is large too, and
our estimate of $10\%$ should stand as a central value of it.
As long as we do not know how to calculate hadronic matrix elements
it will be hard to place a conservative upper bound.

New physics can also be discovered by comparing the two ways to
measure $\gamma$ in the \sm, \ie\ from $b\to c\bar cd$ combined with
$b\to u\bar ud$, and $b\to c\bar us$ combined with $b\to u\bar cs$.
This is not so promising since the
rates are relatively small, and the theoretical uncertainties
are larger. Thus one would require larger effects in order for them to
be observable.
Moreover, isospin
analysis that would substantially reduce the \sm\ uncertainty 
in the $b\to u \bar ud$ would simultaneously remove the 
isospin invariant new 
physics effects from this mode, thus requiring effects in 
the $b\to c\bar us$ mode 
(which were not found in the three models studied here).

To conclude, we have argued in this paper that new physics in the decay
amplitudes of $B$ mesons could lead to observable deviations from the
patterns of $CP$ violation in $B_d$ decays predicted by the \sm. This
is because the small \sm\ uncertainties in these
predictions make even ${\cal O}(5\%)$ effects potentially observable.
This is in contrast to the more commonly studied case of new physics
contributions to the $B_d-\bar B_d$ mixing amplitudes, where the
uncertainty in the \sm\ predictions requires effects of ${\cal O}(1)$
in order to be observable. We gave as examples three models where
measurable effects are allowed.

\acknowledgements
We would like to acknowledge useful discussions with 
G. Buchalla, J. Hewett,
A. Kagan, Z. Ligeti, Y. Nir, M. Peskin, H. Quinn, J. Rosner and J. Wells.

{\tighten

\begin{table}
  \begin{tabular}{c|c|c|c|c|c|c}
\qquad Mode \qquad &  SM angle $(\phi_0)$ & $\dphi_{SM}$ & $ \dphi_A$ & 
$\dphi_B$ & $\dphi_C$ & $BR$\\
\hline 
$b\to c\bar cs$ & $\beta$ & 0 & $ 0.1 $ & $ 0.1 $ & $ 0.1 $ &
 $7 \times 10^{-4}$
\\ 
$b\to c\bar cd$ & $\beta$ & $ 0.1$ & $ 0.2$ &$ 0.6$ &$ 0.6$ & 
 $4 \times 10^{-4}$ \\
$b\to c \bar ud$ & $\beta$ & $ 0.05$ & 0 & 0 & $ 0.5$ &
 $10^{-5}$ \\
$b\to s\bar ss$ & $ \beta $ & 0.04 & $ 1$ & $ 1$ & 0 &
 $10^{-5}$ \\
$b\to u\bar ud$ & $\beta + \gamma$ & $ 0.4$ & $ 0.4$ & 
$ 1$ & 0  &
 $10^{-5}$ \\
$b\to u\bar cs$ &  $\gamma$  & 0 & 0 & 0 & 0  &
 $10^{-6}$ \\
$b \to d \bar s s$ & $ 0 $ & $ 0.3$ & $ 1$ & $ 1$ &
$ 1$  &
 $10^{-6}$ \\
\end{tabular}
\vskip 12pt 
\caption[tbmodels]
{Summary of the useful modes. The ``SM angle'' entry corresponds to
the angle obtained from $B_d$ decays 
assuming one decay amplitude and to first order in the Wolfenstein 
approximation. The angle $\gamma$ in the mode $b\to u\bar cs$ is 
measured 
after combining with the mode $b\to c\bar us$. 
New contributions to the mixing amplitude would shift all the entries
by $\delta_{m_d}$.
$\dphi$ (defined in Eq.~(\ref{aa}))
corresponds to the (absolute value of the) correction to the universality
prediction within each model:
$\dphi_{SM}$ -- \sm, $\dphi_{A}$ -- Effective Supersymmetry, 
$\dphi_{B}$ -- Models with Enhanced Chromomagnetic Dipole Operators and
$\dphi_{C}$ -- Supersymmetry without R-parity. 
$1$ means that the phase can get any value.
The $BR$ is taken from \cite{BrHo} and is an order of magnitude estimate
for one of the exclusive channels that can be used in each inclusive mode.
For the $b \to c \bar u d$ mode 
the $BR$ stands for the product 
$BR(B_d \to \bar D \rho) \times BR(\bar D \to f_{CP})$
where $f_{CP}$ is a $CP$ eigenstate.
}
\label{sumtab2}
\end{table}
}

\end{document}